\documentclass[runningheads,a4paper]{llncs}
\usepackage{graphicx, subfigure}
\usepackage{amsmath}  
\usepackage{array}
\usepackage{algorithm}
\usepackage[noend]{algpseudocode}
\numberwithin{equation}{section}
\usepackage{amssymb}
\usepackage{graphicx}

\begin{document}

\title{Pseudo-Cores: The Terminus of an Intelligent Viral Meme's Trajectory}
%
\titlerunning{Pseudo-Cores and its effects on the Network.} 
%
%
\author{Yayati Gupta\inst{1}, Debarati Das\inst{2}
\and S. R. S. Iyengar\inst{1}}

\authorrunning{Pseudo-Cores and its effects on the Network.}
%
\tocauthor{Yayati Gupta, Debarati Das and S. R. S. Iyengar}
\institute{Indian Institute of Technology Ropar\\
\email{yayati.gupta@iitrpr.ac.in, sudarshan@iitrpr.ac.in}
\and
PES Institute of Technology, Bangalore\\
\email{debarati.d1994@gmail.com }}

\maketitle              

\begin{abstract}

Comprehending the virality of a meme can help scientists address problems pertaining to disciplines like epidemiology and digital marketing. Therefore, it does not come as a surprise that meme virality stands out as an integral component of research in complex networks, today.
In this paper, we explore the possibility of artificially inducing virality in a meme by intelligently directing a meme's trajectory in the network. Keeping in mind the importance of core nodes in a core-periphery structure, we propose two shell-based hill climbing algorithms to determine the path from a node in the periphery shell (where the memes generally originate) to the core of the network. On performing further simulations and analysis on the network’s behavioral characteristics, we were also able to unearth specialized shells which we termed \emph{“Pseudo-Cores”}. These shells emulate the behavior of the core in terms of the size of the cascade triggered. In our experiments, we have considered two sets for the target nodes, one being core and the other being any of the pseudo-cores.  We compare our algorithms against already existing path finding algorithms and validate the better performance of our algorithms experimentally.
\end{abstract}

\section{Introduction}

Korean pop star Psy launched a quirky music video ``Gangnam Style'' in 2012 which received almost 1 billion views on Youtube. In India, the song ``Why this Kolaveri Di'' by Dhanush became an instant social rage in 2011. It earned more than 10,500,000 YouTube views by November 2011. Infact, Dhanush even gave a talk at Indian Institute of Management about how they made their music video viral. Investigating the reason behind the virality of these videos remains a very interesting research question. People contemplated upon their popularity and proposed that the unique dance moves, upbeat tone and the visual nature of the Gangnam style made the video viral \cite{wentgangnam}. Similarly ``Why this Kolaveri di'' gained a cult status because of the distinct ``Tanglish\footnote{portmanteau word of Tamil and English}'' style. Many of the Lady Gaga's videos are viral because of the unique costumes donned by her in them. Now, let us take into consideration a completely contrasting example. On 8 January 2010, Paul ``Bear'' Vasquez uploaded a video of a double rainbow near his home. It was the video of a man simply exclaiming about the double rainbow he witnessed outside his home. What was surprising is that this seemingly mundane video got close to 23 million views in 2010. If the unique intrinsic characteristics of a meme are responsible in making it viral, what made this apparently benign ``Double Rainbow'' video go viral?
While the ``Gangnam Style'' and ``Kolaveri Di'' videos clearly illustrate that intrinsic contagiousness contributes to virality, the example of the ``Double Rainbow'' makes it evident that this can not be the only factor in play for making a meme viral.\\

Kevin Alloca, Youtube's trends manager elegantly addressed this question of meme virality in his TED talk - ``Why Videos go Viral ?''\cite{allocca2011videos}. He identifies three factors which contribute to the virality of Youtube videos, namely \emph{ 1) Tastemakers 2) Community participation} and \emph{3) Complete unexpectedness of the meme(intrinsic contagiousness).} Tastemakers are the people in YouTube who belong to a group having greater reach to rest of the YouTube population. In the case of the ``Double Rainbow'' example quoted above, on 3 July 2010, the popular American television host, Jimmy Kimmel posted about the video on Twitter. He, as a ``tastemaker'' catalysed the popularity of the video granting it over 1 million views. A similar phenomenon was observed with the Rebecca Black's song- ``Friday''. This song was joked about in several blogs but grew truly viral after Michael J Nelson from Mystery Science Theatre posted a joke about it. A group of tastemakers contributed to this song's virality and soon the song became a seed to many parodies. It is evident from these examples that the influential spreaders in a network carry potential to induce virality in passable, non exemplary memes. Kitsak et al. \cite{kitsak2010identification} have put forward an algorithm to identify influential spreaders and have proved that a meme gets viral once it reaches these spreaders. They term such spreaders in the network the core of the network. Consequently, we can infer, that if a non exemplary meme makes its way into the core of a network it could quite easily become viral.\\

Taking all the aforementioned facts in account, we propose algorithms that can help a user intelligently guide a meme towards the core of a network. So, if the information originates in the periphery of a network, our problem can be reduced to a path finding algorithm, given that the source is a node in the periphery of the network and destination is the core. Considering the core (the innermost shell of the network) as the epicenter of a contagion spreading in the network, we probe upon the question: \emph{"How can we intelligently target links in the network to quickly reach the core?"}.  The main contributions of the paper are as follows:
\begin{enumerate}
\item Evaluation of different properties of the shells provided by the k-shell decomposition algorithm.
\item Employment of the properties of these shells to take an intelligent walk towards the destination. This is also the key idea of the paper i.e. to utilise the presence of multiple shells in a network to effectively reach its core starting from the periphery. 
\item Unearthing specialized shells which mimic the core and help a meme go viral. We call these shells as \emph{``Pseudo-Cores''}. 
\end{enumerate}
These revelations of our experiments can have a tremendous impact on the fields of information propagation as well as epidemiology. Not only can intelligent pathways be formulated to propagate information, but also if such pathways could be simulated, preventive checkpoints could be placed appropriately to halt infection spread.

\subsection{Core : The Destination for a Viral Meme}

In our proposal, we have visualized the social network as having a meso scale characteristic called as core-periphery structure. The notion of a core-periphery structure was defined by  Borgatti and Everett in their seminal paper \cite{borgatti2000models} in 2000. The core was defined as a set of nodes densely connected to each other having a large number of connections to the periphery nodes. On the other hand, the periphery nodes although connected to the core nodes are largely disconnected amongst themselves. It has been proved that most networks existing in the nature possess core-periphery structure \cite{Zafarani+Liu:2009} \footnote{Most of the networks in real world are scale free \cite{barabasi2009scale}. Works done by Della et al. \cite{della2013profiling} and Liu et al. \cite{Zafarani+Liu:2009} prove that scale-free networks usually possess a core-periphery structure. Therefore by transition, it becomes evident that a social network is a scale free network as well as a core-periphery structure.}. This result has far reaching consequence on meme virality.\\

The application of core-periphery structure came to light as a consequence of the criticism of the Milgram's experiment \cite{travers1969experimental} by Judith Kleinfield \cite{kleinfeld2002small}. Milgram's small world experiment aimed at passing a letter from a source person to the target person through a chain of acquaintances. The letter could only be forwarded to a person whom the participant knew on a first name basis. Milgram's experiment found the median of the chain lengths over multiple experiments to be six. Later on, Kleinfield observed that the success rates of the recreations of Milgram's experiment were very low \cite{kleinfeld2002could}. Kleinberg clarified that the chain length was observed to be six only when the destination was a high status individual or ``core'', otherwise the chain length was higher \cite{easley2010networks}. In their work, Kleinberg et al. proved that high status people have access to higher privileges as compared to the common crowd. These high status individuals can travel widely and be a part of many clubs and sub-networks making connections over social boundaries. This makes these people well connected amongst themselves as well as with the rest of the network. As this elite group has a higher reach over the network, they are more easily accessible by the low status people in the network. Kitsak et al. \cite{kitsak2010identification} have also shown that the influential spreaders in a network are those which lie at its core and that once any meme reaches the core of a network, its chances to go viral increases drastically. These two properties of maximum spreading power and high reachability make the core a suitable destination for a viral meme.

\subsection{K-Shell Decomposition}

This algorithm was proposed by Kitsak et al. \cite{kitsak2010identification} in order to determine the core in a network. The algorithm works by recursively pruning the lower degree nodes in a network. First the nodes having $degree$ $d(u) \leq 1$ are pruned recursively till there
 are no nodes of degree 1 left in the network. Similarly, the process is repeated for the nodes $k$ having degree $d(k)$ where $d(k)$=2, 3, ....., n-1 till the graph becomes empty. At every step, the pruned nodes can be visualised as being kept in a basket. The baskets formed earlier represent the least core nodes. The nodes in one basket form one shell of the graph and a higher shell number represents higher coreness. One of the key facts of the algorithm is that the nodes having high degrees lying at the periphery are pruned earlier because of recursive pruning. Therefore, it is not necessary that a person having high degree should also be a core node.

\section{Experimental Observations Inciting the Algorithm}

In this section, we explain some of the experiments we performed and the observations that led us to the proposed idea. K-Shell decomposition \cite{kitsak2010identification} is one of the most popular algorithms to decompose a network into multiple shells of influence. We decompose a network employing k-shell decomposition algorithm and observe the properties of various shells. In this decomposed network, the innermost shell, or the core, is the most dense subgraph having the highest closeness centrality. As we move outward towards other shells, density and closeness both decrease and the outermost shells are called the periphery shells.
Core-periphery structures have been studied greatly ever since they were introduced in 2000. However the research attention was focused mostly on the core, and scientists have not yet tried to harness the power of the intermediary shells in the network. In this paper we investigate the properties of all the shells in a network and then observe how the network structure in these shells can be efficiently used to guide the meme in a correct direction and make it viral.\\

\subsection{Experiments investigating the correlation of network properties with the shell number}

\subsubsection{Distribution of nodes across shells:}
We observe the number of nodes present in every shell of the network as we move inwards from the periphery to the core. While moving inside, we intuitively expected this number to strictly decrease, but the plot in figure 1 indicates something unexpected.
It is seen that the number of nodes do not necessarily follow an ideal rectangular hyperbola curve fitting the equation $y=\frac{1}{x}$.
The plot shows increase in the population of nodes in the intermediary shells as well. But it can be observed in most of the networks that the core is the shell with the least diameter.
\begin{figure}[h!]
\centering
\includegraphics[width=9 cm]{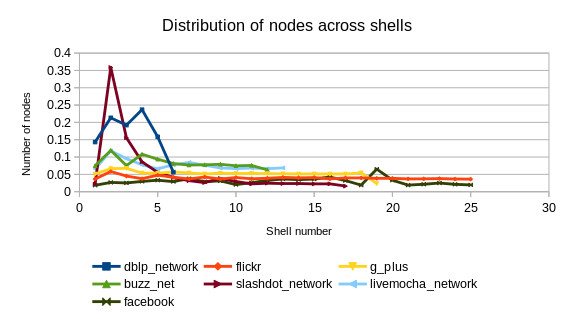}
\label{numnodes}
\caption{Distribution of nodes across shells for various networks}
\end{figure}

\subsubsection{Density of shells:}
\begin{figure}[h!]
    \centering
    \subfigure[Density of shells in buzznet, slashdot and livemocha]
    {
       \includegraphics[width=5.8 cm]{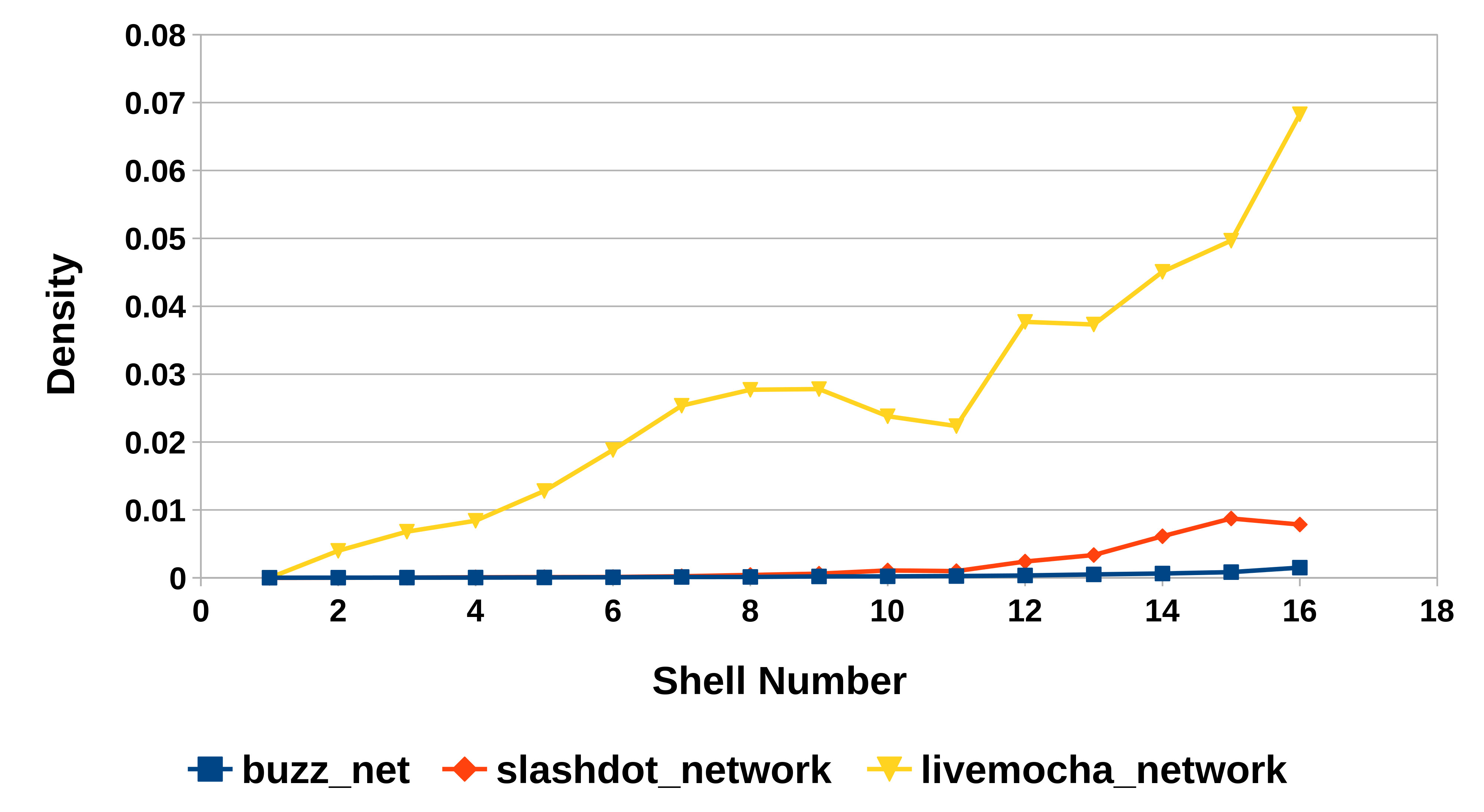}
       
    }
    \subfigure[Density of shells in DBLP]
    {
       \includegraphics[width=5.8 cm]{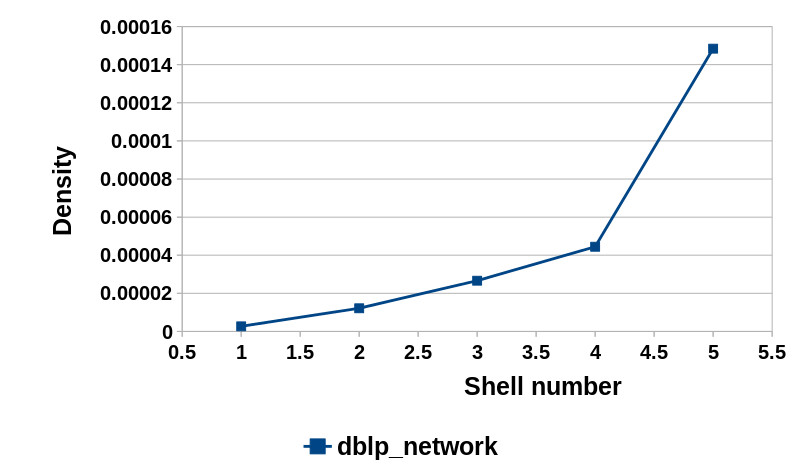}
       
    }    \subfigure[Density of shells in Flickr, Google Plus and Facebook]
    {
       \includegraphics[width=5.8 cm]{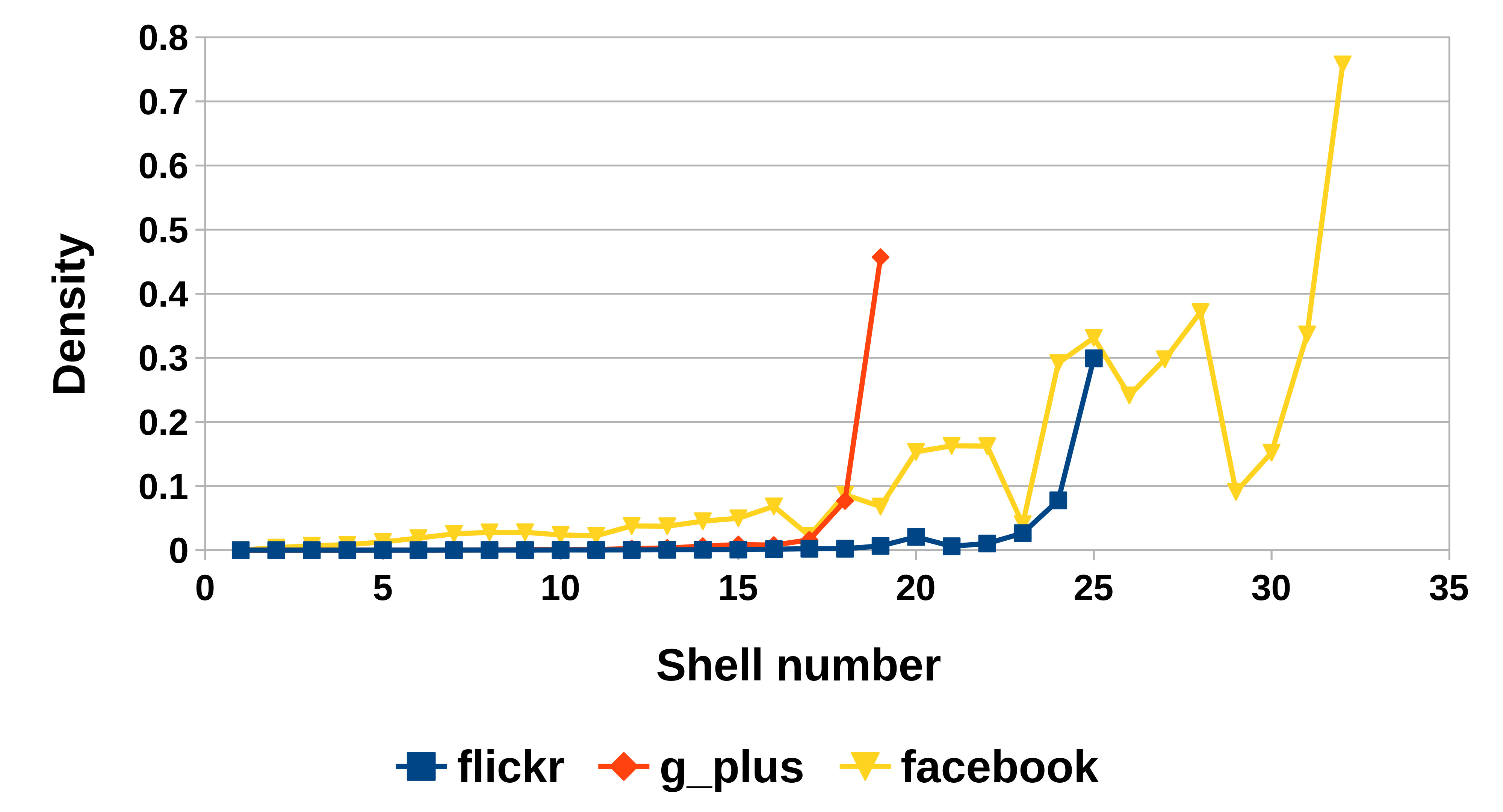}
       
    }
        \caption{Shellwise Density distribution in Real world networks}
        \label{density}
\end{figure}

Let us represent the given network by a graph $G(V,E)$. The density of shell $i$ is defined as the density of the induced subgraph $S_i$ on the vertices that lie in shell $i$ of the network $G$. Let $|V(S_i)|$ be the number of nodes in the subgraph $S_i$ and $|E(S_i)|$ be the number of edges. Then, the density of shell $i$ can be given as $\frac{|E(S_i)|}{\binom {|V(S_i)|}{2}}$.
As observed from figure \ref{density}, Buzznet, Slashdot, Livemocha, Flickr and Google Plus are some social networks which show a uniformly exponentially increasing curve of density vs. shell number. A similar curve is observed in the case of collaboration network DBLP. When the core shell is encountered in these networks, there is a sudden but powerful spike in density. This is probably one of the contributory factors to the spreading power of the core.
However, in the case of social networks like Facebook, the curve does not accelerate only on reaching the core, there are quite a few spikes in between and the curve does not remain monotonically increasing. However, even in this case, core is observed to possess the maximum density.
As a result we can safely assume that the high density of the core is one of the reason for the high spreading power of the core. The plots are shown in figure \ref{density}.

\subsubsection{Cascading Power of Different Shells in a Network:}
\begin{figure}[h!]
    \centering
    \subfigure[Cascading power of facebook shells]
    {
       \includegraphics[width=5.8 cm]{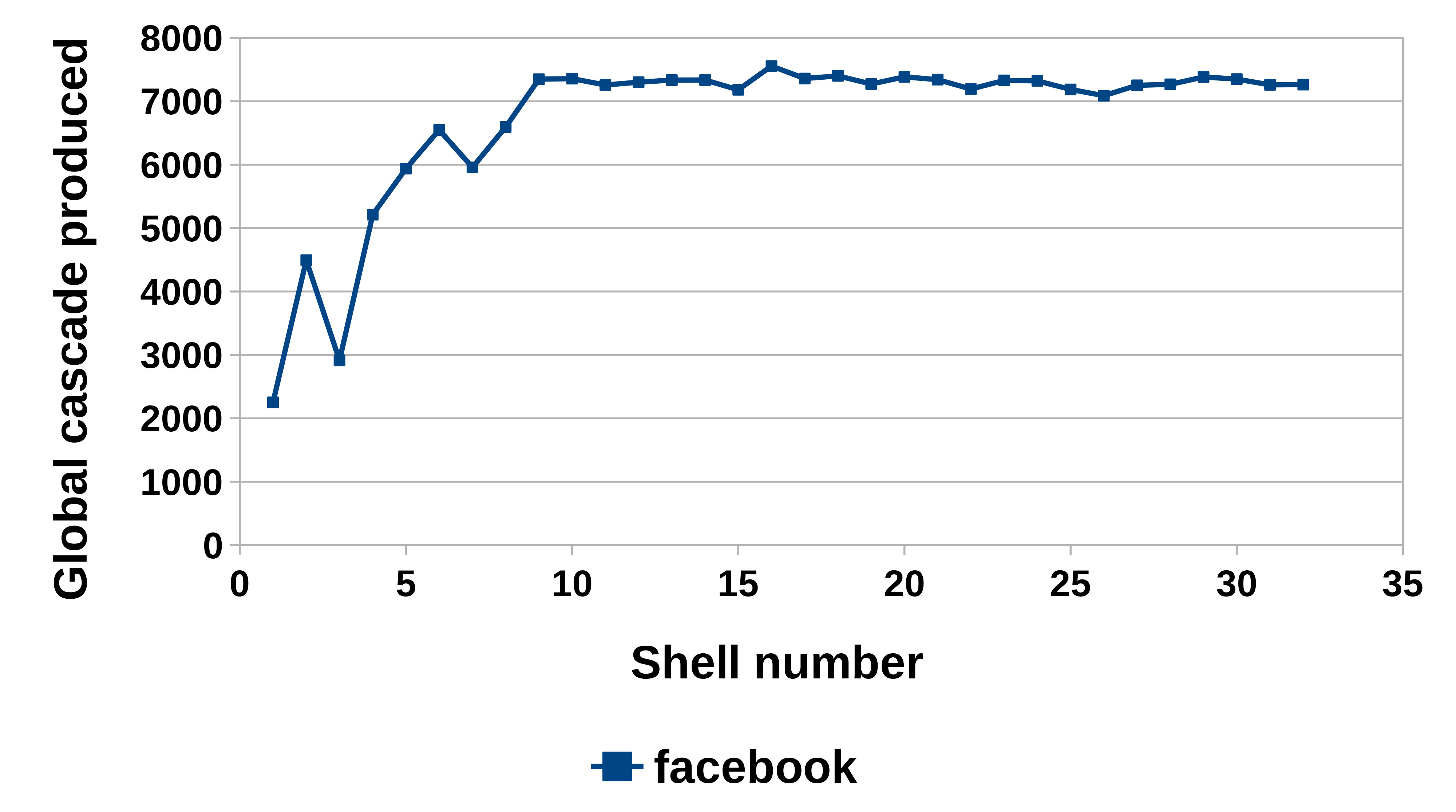}
       
    }
    \subfigure[Cascading power of other real world network's shells]
    {
       \includegraphics[width=5.8 cm]{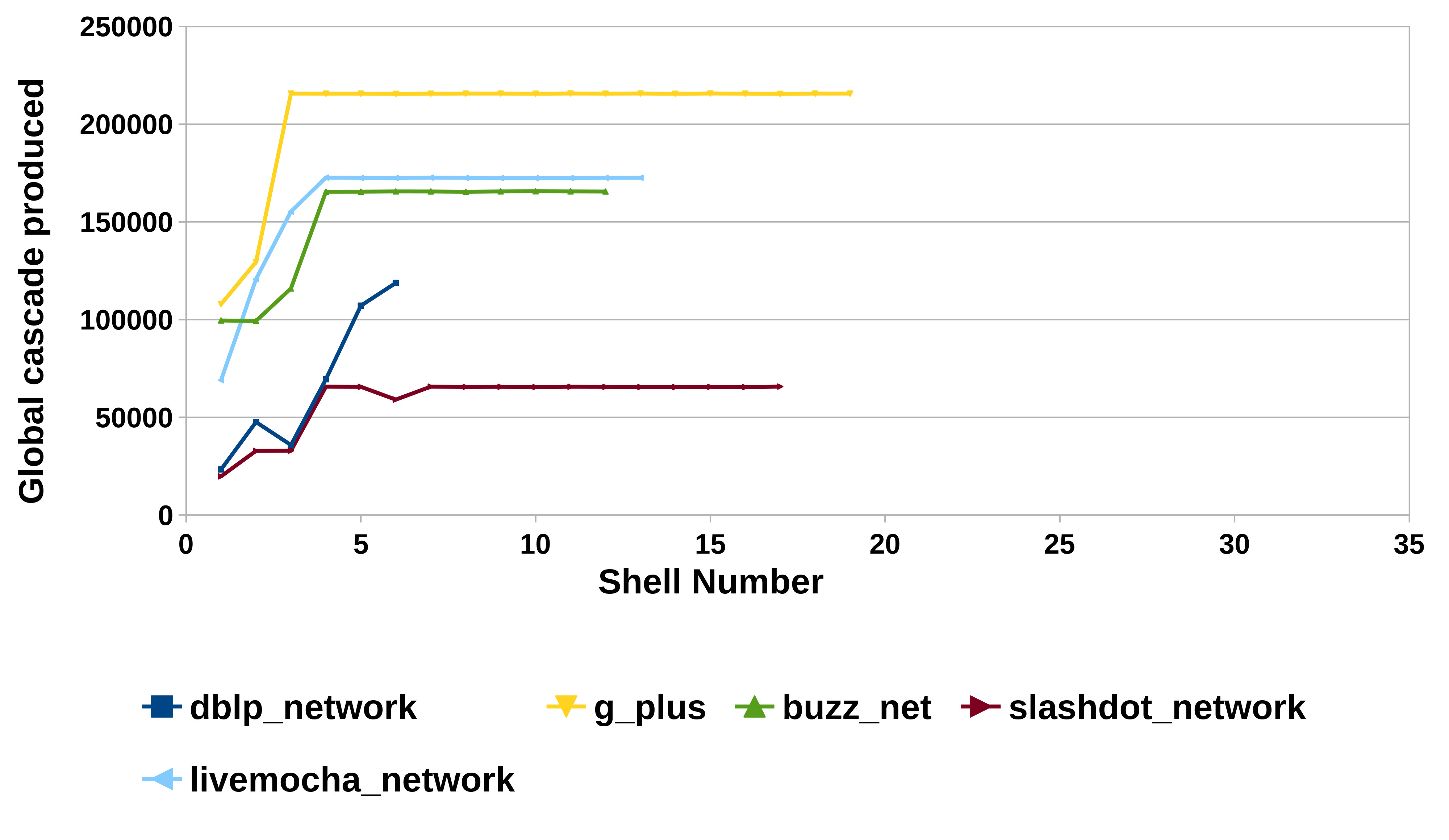}
       
    }
  \caption{Shell wise cascading power distribution}
  \label{cp}
\end{figure}

We choose nodes uniformly at random from a shell. Then we use the independent cascade model for meme propagation and initiate the process of infection over this network. Therefore at every iteration, each infected node infects its neighbours with a constant probability. We note that every node gets only one chance to infect its neighbours. The process stops as soon as it reaches an iteration where no new node is infected. The number of nodes infected by the end of the process is called the cascading power of a shell. 
The plots in figure \ref{cp} represent the meme cascade size produced if the cascade starts from some of the nodes of a particular shell. One ideally expects the cascade to accelerate when core shell is encountered, but it is observed that the acceleration point is reached much before the point where the meme reaches the core.
This intriguing fact led us to investigate the existence of a ``pseudo- core shell'' or a shell which provides something akin to an escape velocity for the meme to become viral i.e. once this shell has been visited in the meme trajectory, the meme goes viral and there is no longer any need for it to target some higher shell node. This hypothesis has many large scale implications. For example- If I were a political analyst and I were trying to find which person to infect in a political network, I would no longer have to infect the most influential politicians or relatively insulated core nodes. Infecting someone relatively less influential(if I could find that this person lies in a pseudo-core shell) would cause the same effect. \\
We compare a set of path finding algorithms in the next section and further observe if changing the destination to pseudo-core has any relative impact on the time taken for the path.

\section{Algorithms}
We describe algorithms in this section to find a path from the periphery of a network to the destination, which we initially assume as core. Later we apply the same algorithm with pseudo-cores as the destination and report the improved results. We describe two already implemented algorithms in table \ref{existing}.

\begin{table}
\begin{center}
\begin{tabular}{ | m{5cm} | m{7cm}| }
\hline
\textbf{Random walk algorithm\cite{noh2004random}} & \textbf{Degree based hill climbing \cite{adamic2001search}} \\
\hline
This algorithm involves a node inspecting its neighbours at every step and selecting one of them randomly. If the chosen neighbour is a core node, the algorithm terminates, else the selection of the random neighbours continues. Random walk algorithm (without repetition of nodes) has a time complexity of $O(n)$. &
This algorithm uses a hill climbing approach based on the degree of the nodes in the network. At every step, a node looks at its neighbours and chooses the unexplored node having the highest degree. If the chosen node is a core node, the algorithm terminates, else the process continues. As hill climbing algorithms have a complexity of $O(n)$ where n is the number of nodes and finding degree of all nodes takes $O(m)$ time, degree based hill Climbing has a time complexity of max$[O(n),O(m)] \sim O(n)$ in sparse graphs. \\
\hline
\end{tabular}
\end{center}
\caption{Existing algorithms}
\label{existing}
\end{table}

Below, we propose two hill climbing algorithms based on the shell numbers of the nodes. After the network has been decomposed into shells, the entire system can be visualized as a circular maze made up of concentric circles(shells). The goal of the algorithm is to intelligently move from the outermost shell to the innermost shell. There are inter-shell edges that help a user in taking such a walk across shells, while the intra-shell edges help the user to traverse a shell.

\subsubsection{Algorithm 1- Shell based Hill Climbing Approach (SH):}

Let $G(V,E)$ represent the graph where $V(G)$ is the set of vertices and $E(G)$ is the set of edges. Let the number of vertices and edges in G be n and m respectively. $shell(u)$ represents the shell number of a node $u$ as calculated by the k-shell decomposition algorithm. $start$ is the periphery node from where the meme starts spreading. $N_G{(u)}$ represents the set of neighbours of node $u$ in the graph $G$. The proposed SH approach has a complexity of $max[O(m+n),O(n)] \sim O(n)$ in sparse graphs.

\begin{algorithm}
\caption{Shell Based Hill Climbing(SH)}\label{euclid}
\begin{algorithmic}
\Procedure{FindNumsteps}{}
\State \textbf{Input}:- Graph $G(V,E)$, Starting node $start$
\State \textbf{Output}: Number of steps taken by the algorithm to terminate
\State Apply k-shell decomposition and calculate $shell(u)\ \forall u \in V(G)$
\State $visited[u] \gets`false'\ \forall u \in V(G)$
\State $numsteps\ \gets 0$
\State $current \gets start$
\State $visited[current] \gets `true'$
\While{$current$ is not a core node}
\State $v_1\ \gets argmax_{u \in N_G(current) \wedge visited[u] = `false'} shell(u)$
\If{$shell(v_1) \leq shell(current)$}
\State $v_2 \gets random\ node\ u\ \in N_G(current) \wedge visited[u] = `false' $
\State $current \gets v_2$
\Else
\State $current \gets v_1$
\EndIf
\State $numsteps \gets numsteps+1$
\EndWhile
\State return $numsteps$
\EndProcedure
\end{algorithmic}
\end{algorithm}

\subsubsection{Algorithm 2 - Intershell Hill Climbing with Intrashell Degree Based Approach(SA):}

Algorithm 2 is a modification of algorithm 1 and utilises the idea that a node with very high degree will cover most of the shell. If this node is chosen, it would greatly reduce the number of steps required to traverse a shell. Let the number of vertices and edges in G be n and m respectively. $N_G{(u)}$ represents the set of neighbours of node $u$ in the graph $G$. The proposed SA approach has a complexity of $max[O(m+n),O(n),O(m)] \sim O(n)$ in sparse graphs.

\begin{algorithm}
\caption{Improved Shell Based Hill Climbing(SA)}\label{euclid}
\begin{algorithmic}[1]
\Procedure{FindNumsteps}{}
\State \textbf{Input}:- Graph $G(V,E)$, Starting node $start$
\State \textbf{Output}: Number of steps taken by the algorithm to terminate
\State Apply k-shell decomposition and calculate $shell(u)\ \forall u \in V(G)$
\State $visited[u] \gets \lq false\rq\ \forall u \in V(G)$
\State $numsteps\ \gets 0$
\State $current \gets start$
\State $visited[current] \gets `true'$
\While{$current$ is not a core node}
\State $v_1\ \gets argmax_{u \in N_G(current) \wedge visited[u] = `false'} shell(u)$
\If{$shell(v_1) \leq shell(current)$}
\State $v_2  \gets argmax_{u \in N_G(current) \wedge visited[u] = `false'} degree(u)$
\State $current \gets v_2$
\Else
\State $current \gets v_1$
\EndIf
\State $numsteps \gets numsteps+1$
\EndWhile
\State return $numsteps$
\EndProcedure
\end{algorithmic}
\end{algorithm}

\begin{figure}[h!]
    \centering
    \subfigure[Algorithm1- Shell Based Hill Climbing(SH)]
    {
       \includegraphics[width=5.8 cm]{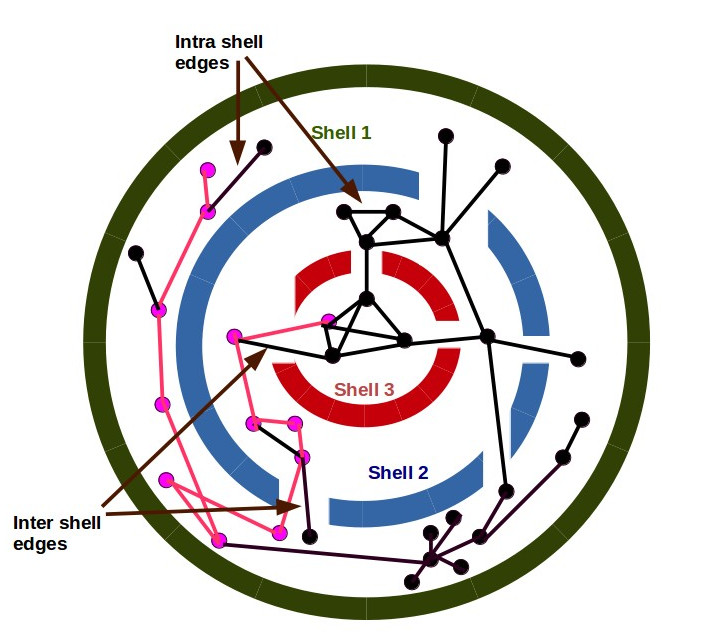}
       
    }
    \subfigure[Algorithm2- Modified Shell Based Hill Climbing(SA)]
    {
       \includegraphics[width=5.8 cm]{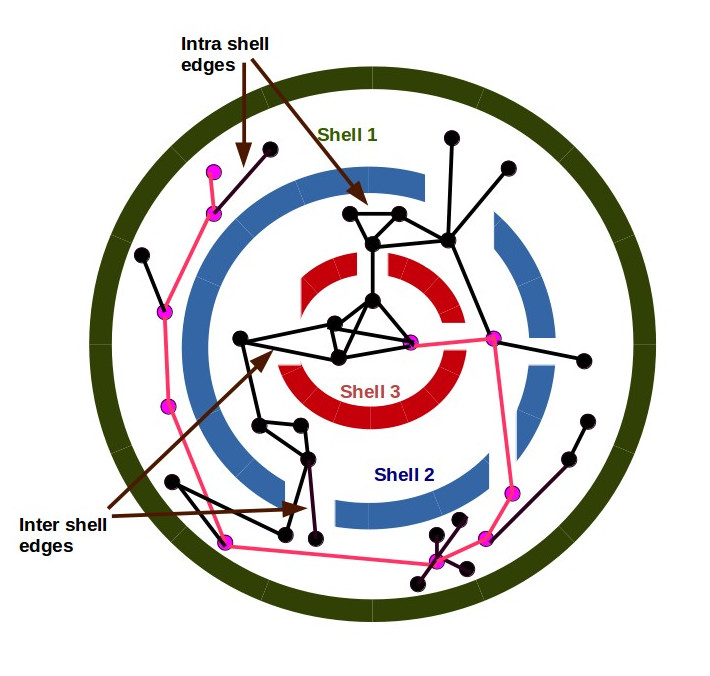}
       
    }
  \caption{Proposed Algorithms: The path denoted in the pink edges is the path chosen by the corresponding algorithm to move towards the core.}
  \label{algo}
\end{figure}

\section{Experimental Results}
To evaluate the performance of the algorithms mentioned in the above section, we select periphery nodes from shell 1(periphery) in a network and for each of these nodes, we find the number of steps taken to reach the core. We term each run from a periphery node as an instance of the problem. Therefore, we can say the number of instances is equal to the number of periphery nodes.
It is observed that more than 80\% of the walks conclude in a maximum of 15 steps in most of the datasets. Without the loss of generality, we have ignored the trivial case where source nodes are directly connected to the core as the path length in these cases is 1.

Let $R$ be a random variable whose value ranges from $2$ to $k$. $R$ depicts the number of steps taken by the algorithm to terminate.
Let $P(R=k)$ be the Probability of value of $R$ being $k$
where $k$ ranges from $2$ to $15$.
We plot the cumulative probability distribution function of R. X axis indicates all possible values of R while Y axis shows the probability of $R \leq k$.

The plots given below validate that the proposed algorithms cover most of the instances in very less number of steps as compared to the existing path finding algorithms. The highest line in the curve represents the most efficient algorithm. In the case of Facebook network, the proposed algorithms cover 80\% of the instances in less than 100 steps. Degree based hill climbing requires around 200 steps to cover 80\% of the instances. In the case of Google Plus, all the three hill climbing based approaches cover 90\% of the instances in less than 3 steps. The results for the rest of the networks are included in Appendix.
In all the cases, the algorithms proposed reach their peak at the earliest proving that they are more optimal with respect to the time taken to reach the destination.The random walk algorithm clearly performs the worst.

\begin{figure}[h!]
    \centering
    \subfigure[Random Walk]
    {
       \includegraphics[width=5.8 cm]{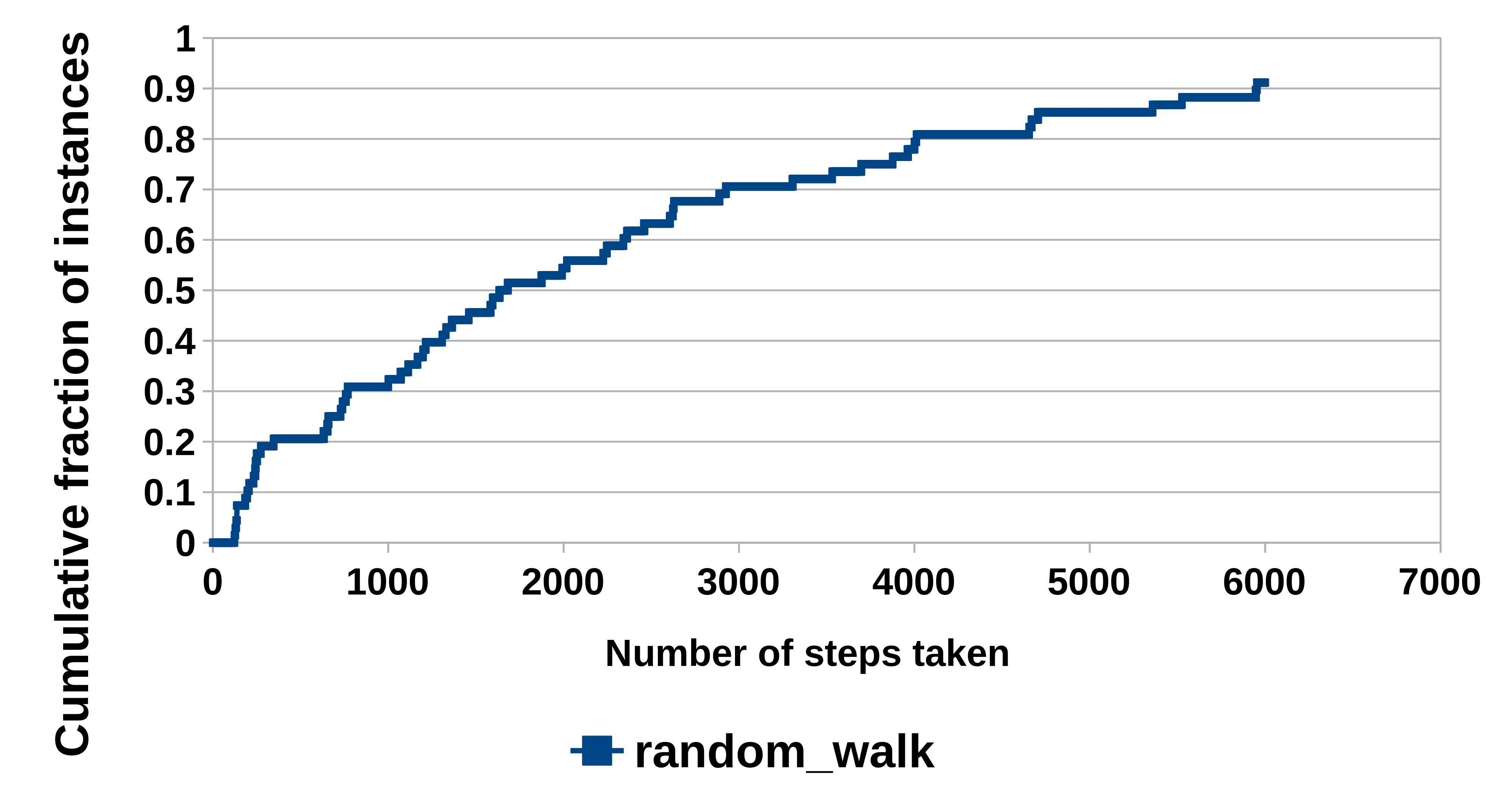}
       
    }
    \subfigure[Shell Based Hill Climbing Algorithms]
    {
       \includegraphics[width=5.8 cm]{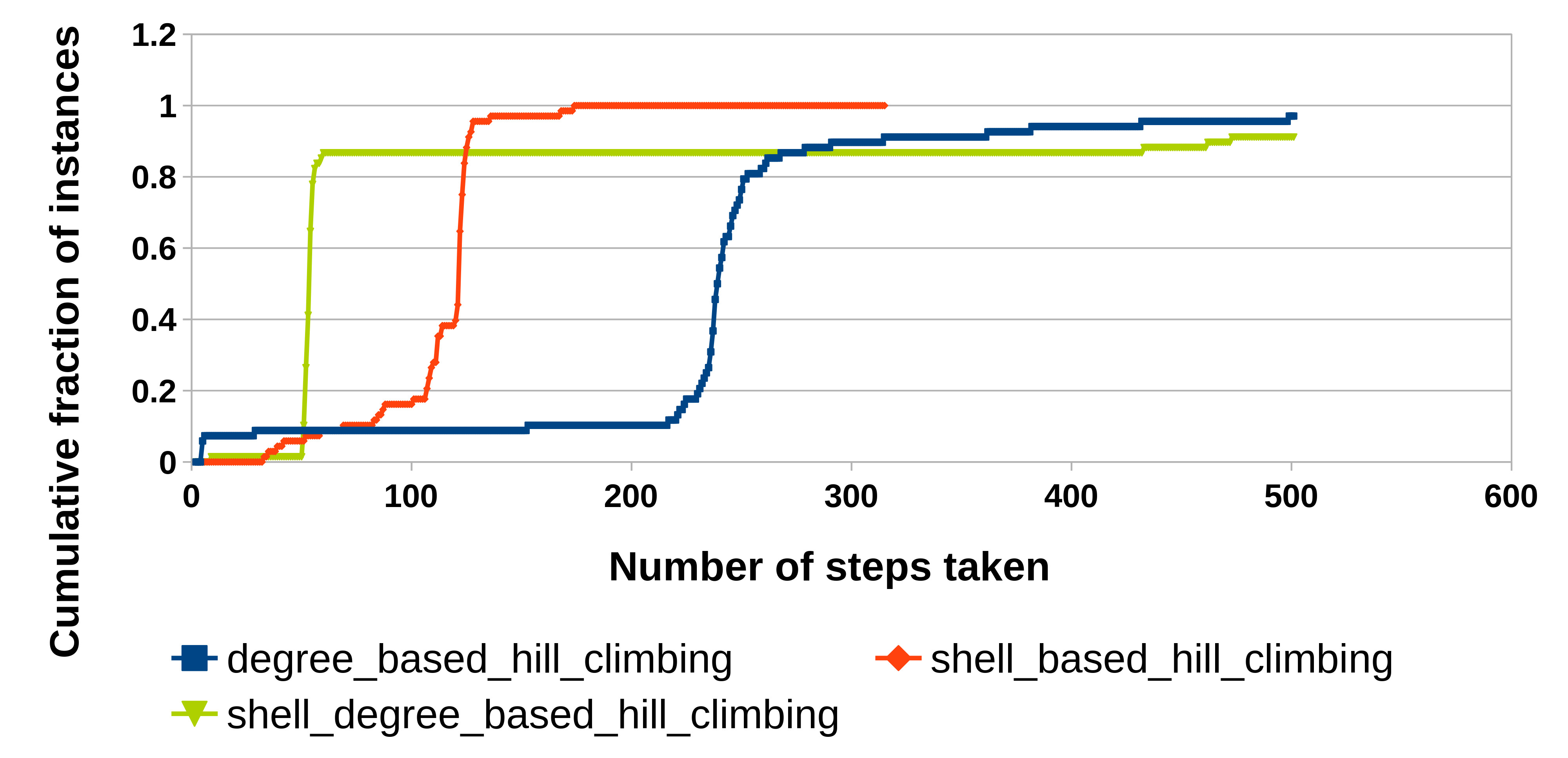}
       
    }
  \caption{Comparison of algorithms for Facebook}
  \label{fb}
\end{figure}

\begin{figure}[h!]
    \centering
    \subfigure[Random Walk]
    {
       \includegraphics[width=5.8 cm]{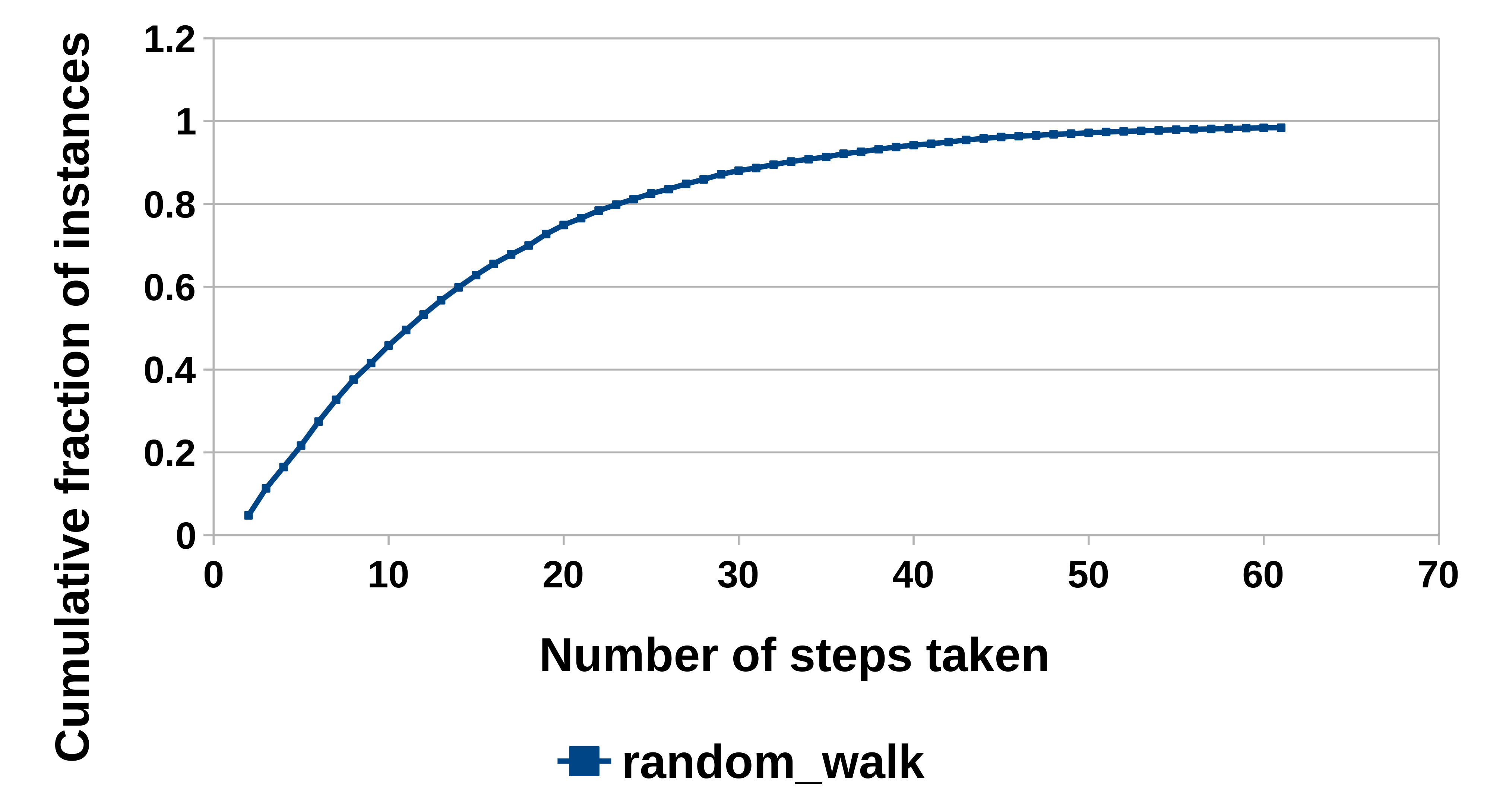}
       
    }
    \subfigure[Shell Based Hill Climbing Algorithms]
    {
       \includegraphics[width=5.8 cm]{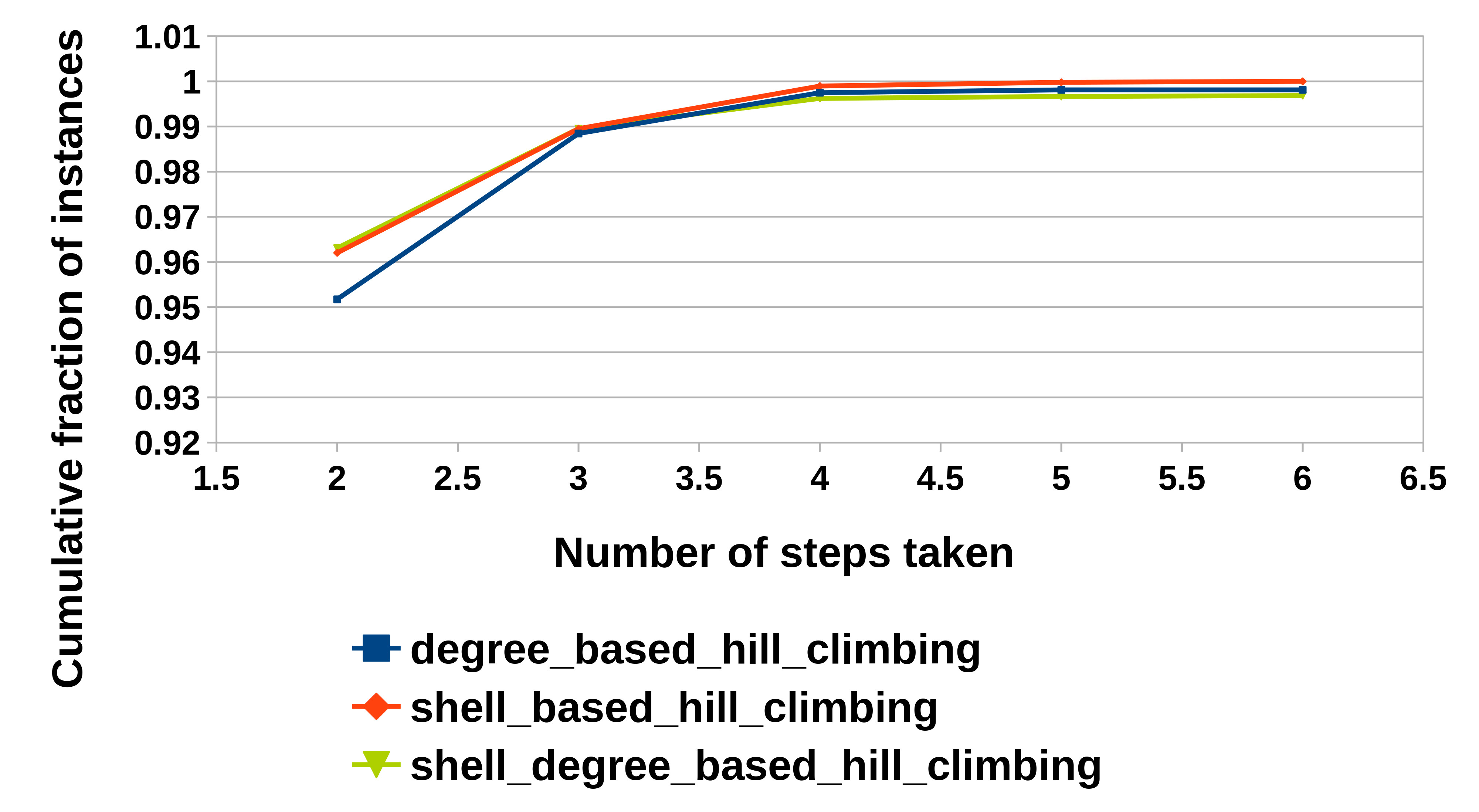}
       
    }
  \caption{Comparison of algorithms for Google Plus}
  \label{gp}
\end{figure}

\begin{figure}[h!]
    \centering
    \subfigure[Facebook]
    {
       \includegraphics[width=5.8 cm]{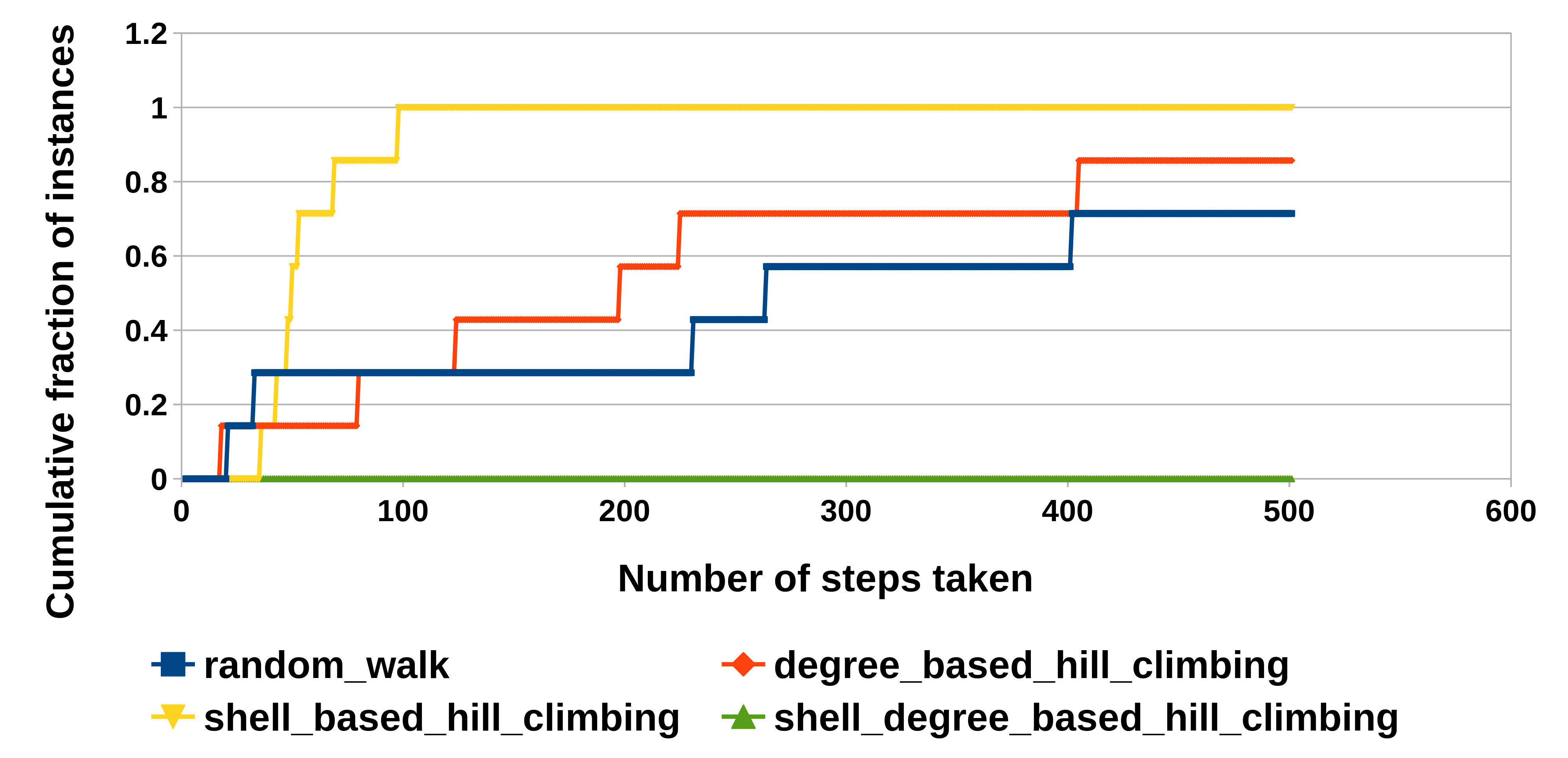}
       
    }
    \subfigure[Google Plus]
    {
       \includegraphics[width=5.8 cm]{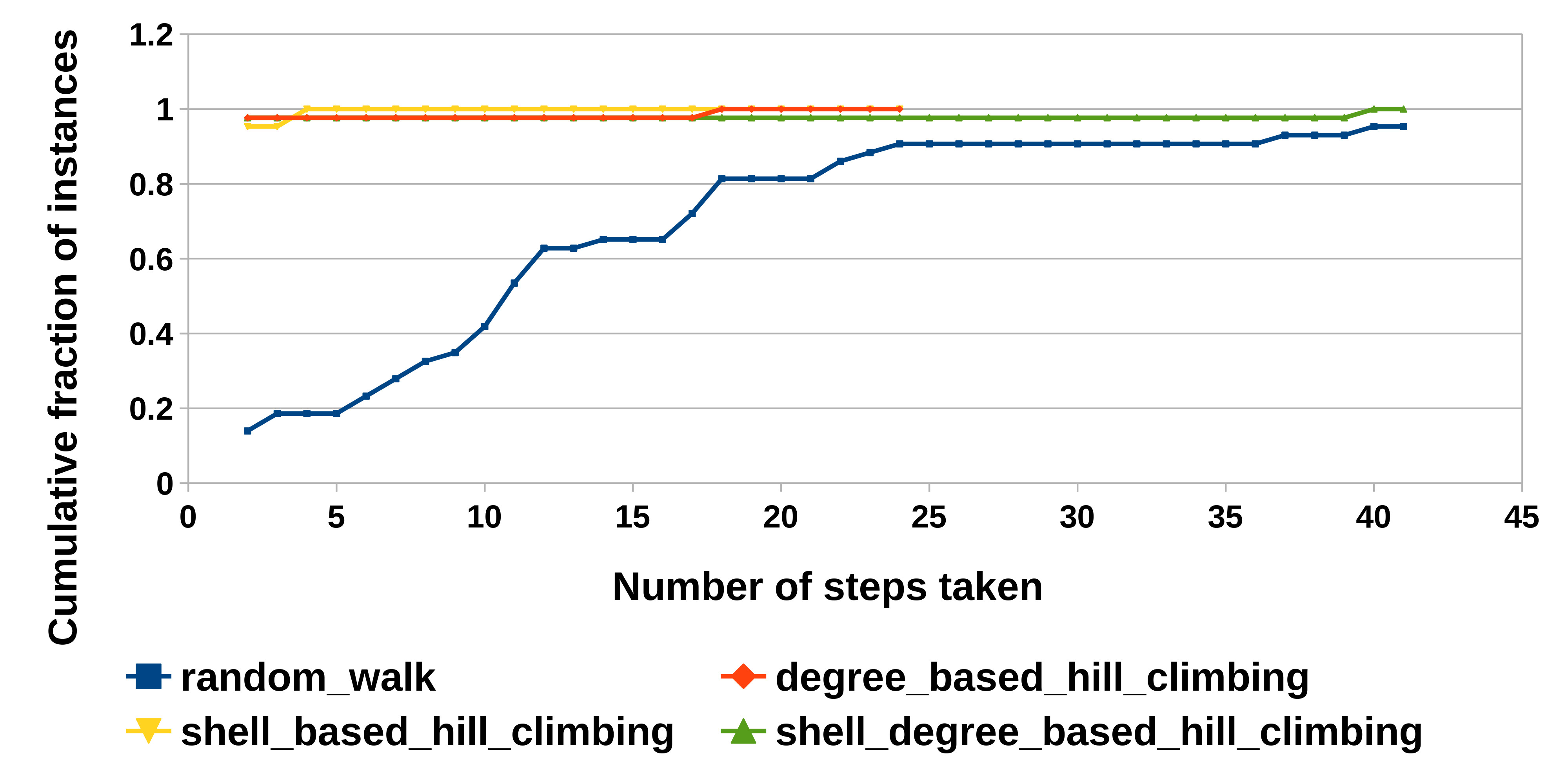}
       
    }
    \subfigure[Slashdot]
    {
       \includegraphics[width=5.8 cm]{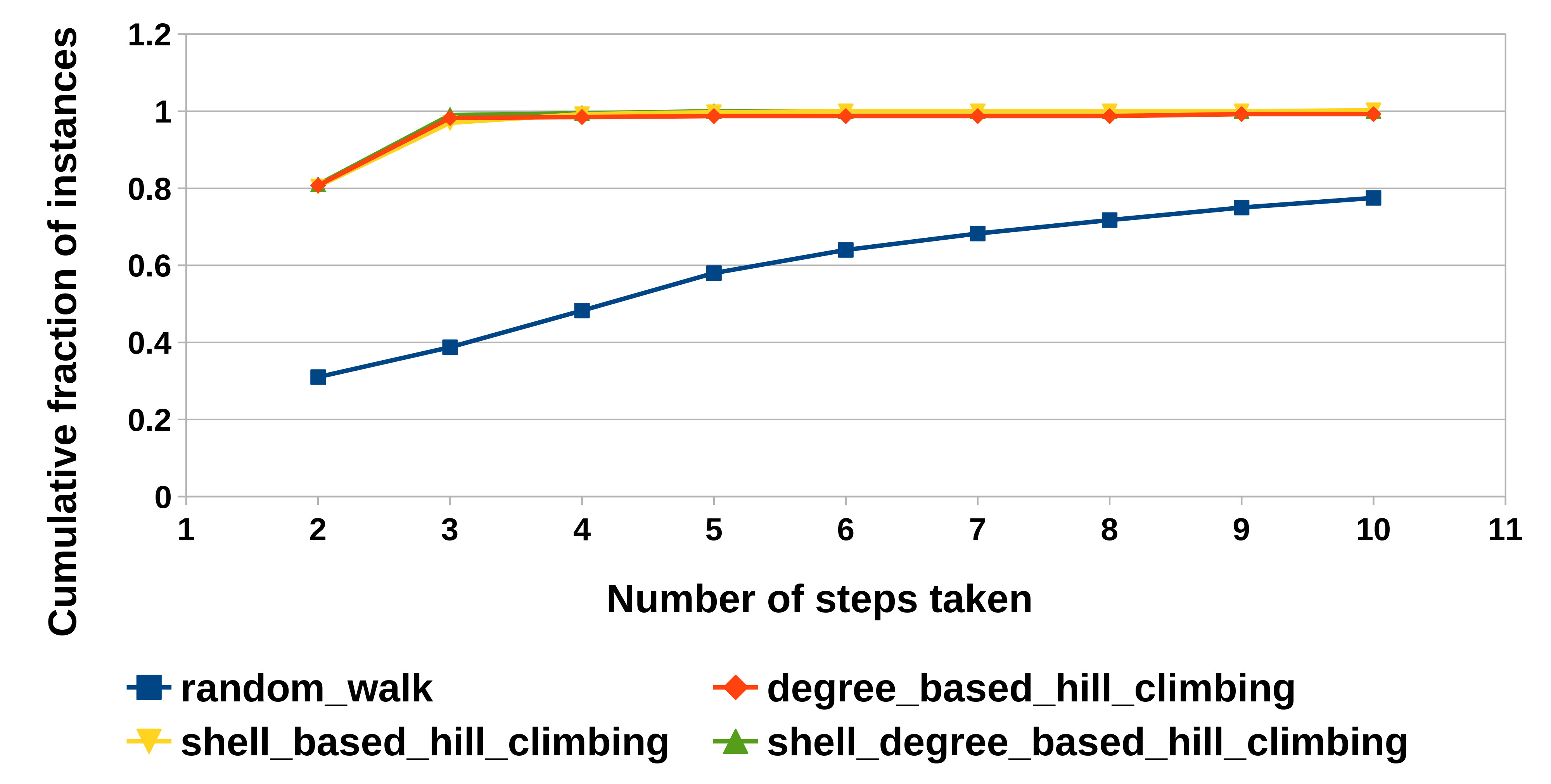}
       
    }
  \caption{Comparison of algorithms for infecting pseudo-cores}
  \label{pseudo}
\end{figure}

Next, we modify the destination to be the pseudo-core shells and observe the cumulative frequency distribution of R as given in figure \ref{pseudo}. In this case also, our proposed algorithms perform better than the other algorithms. Interestingly, the performance of even the random walk algorithm increases drastically when the target is changed to be the pseudo-cores. This indicates that the virality which seems frequent and random in our social as well as biological networks may be because of the presence of pseudo-cores in the network. It is fairly intuitive that it is difficult to target an insulated and well connected core node but it would be relatively simple to hit the pseudo-core and this could be one of the possible explanations for meme virality in a network.
The results for these simulations are shown in figure \ref{pseudo}.

\section{Related Work }

The information derived from the internet is being harnessed in a myriad of applications today. Culotta et al. \cite{culotta2010towards} have used the information potential of a social network to predict epidemics in a population. Social networks act as reservoirs of data which can be used to predict the results of elections \cite{sang2012predicting} as well as patterns in crime \cite{gerber2014predicting}. Meme is a term used to describe a unit of information traversing in a network. These memes behave like biological viruses and evolve over time as suggested by Daley et al. in their work \cite{daley1964epidemics}. Memetics or the study of memes has a wide range of applications in several research areas like Digital Marketing and Epidemiology. This is not surprising as deciphering patterns in any kind of data or trajectories in information flow in the network can have wide range impacts. If for some reason an information goes “viral” - impacts a large portion of the network, then the meme holds more potential in the network for analysis.\\ 

Many approaches have been employed to understand the cause of meme virality. Berger \& Milkman \cite{berger2012makes} employed the content of a meme to predict its virality while Weng et al. observed the similarity between a simple contagion and a viral meme \cite{weng2013virality}. The existence of communities and core-periphery structure \cite{borgatti2000models} are two major discoveries with respect to complex network structure. We applied the studies on complex network structures to understand meme spread and probed upon the question :\emph{ ``Can we intelligently alter the path of a meme flowing through a network to make it go viral?''}.\\

Milgrams experiment\cite{milgram1967small} had a similar aim to find the shortest path from a source person to the target person. For this experiment, breadth first search approach is not suitable as it would lead to flooding of letters in the network. We cannot also assume that a person will advertise a product to each of his/her neighbours. Similarly a DFS might result in several paths which are not optimal. However, It was observed that even though people did not possess an overview of the entire network, they were still able to trace the average 6-hop path between two individuals. This spawned the idea of a decentralised search approach \cite{kleinberg2006complex} ,or the ``Myopic Search'' approach. This greedy heuristic aimed at providing a path from a source to a destination exploring only one new node per iteration which is nearest to the target. We were inspired by this decentralised approach to propose a method to direct a meme in an optimal direction. Our work differentiates from the decentralised algorithms in two ways:
\begin{enumerate}
\item Instead of focusing on one target node, we are trying to attack exactly one node in a group of nodes, termed as core.
\item We are proposing the algorithm for real world networks instead of very well defined lattice like structures.
\end{enumerate}
In this paper, we propose a hill climbing technique by virtue of which a user needs to focus only on one neighbour who gives him/her more benefit as compared to distributing his/her efforts among all the neighbours. Kleinberg et al. \cite{kleinberg2006complex} attempted targeting the optimal selection of seed nodes \cite{kempe2003maximizing}, but as yet no work has been done to identify the optimum destination nodes in information propagation, which is what we are attempting empirically. Though work has been done on path finding algorithms in a social network, we are unaware of any approaches which have combined the idea of visualizing a network as a core-periphery structure and utilizing this property to devise a path finding algorithm over the network. Our work is the first of its kind to the best of our knowledge.

\section{Conclusion}
The paper unravels the potential of a core-periphery structure in imparting virality to a passable/non-exemplary meme. We empirically observed that the innermost core shell has the greatest tendency to trigger a global cascade in the network, thereby increasing the necessity to infect the core quickly in order to cause virality. We have proposed two shell based hill climbing approaches that help a meme to pave an intelligent path to the core, when it originates in the periphery. One of the most important contributions of the paper is the unveiling of the concept of \emph{``Pseudo Core''} shells that have the same cascading impact on the network as a core shell. Intelligently hitting the pseudo-core shell achieves the same virality as that achieved by core shell. As a result the path taken during the trajectory of a viral meme can be reduced. These revelations introduced by our experiments may prove to have huge impact across several disciplines.

\section{Future Work}
\begin{figure}[h!]
\centering
\includegraphics[width=9 cm]{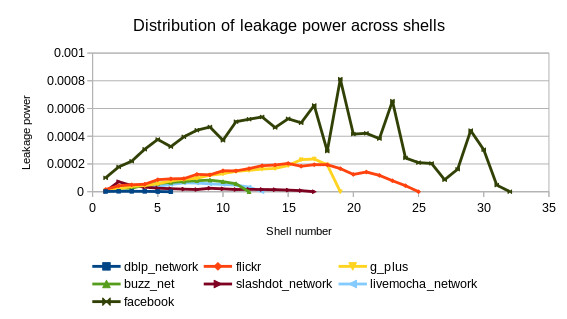}
\label{leakage}
\caption{Distribution of leakage power across shells for various networks}
\end{figure}
While performing experiments to analyse network property effects on shells, we defined a shell parameter which we deem \emph{``Leakage Power''}. Leakage power denotes a shell's potential to take quick and long jumps to the higher numbered shells. We plotted leakage power against shell number and observed that the leakage power was not necessarily the highest in the case of core. There were indications of high leakage power in intermediary shells as well. This is shown in figure \ref{leakage}. This led us to investigate the ideas of \emph{``teleportation shells''} and \emph{``barricade shells''}. The shells having higher leakage powers act as the teleportation shells and can trigger a meme to take longer jumps on its path to the core. On the other hand, the shells having low leakage powers may tend to block a meme inside it, hence suggesting why some memes are non viral. Based on these observations, the algorithms may be altered to provide better results and most importantly answer the bigger question which is \emph{``Why do some memes selectively go viral ?''}. Another interesting research idea is to come up with an approximation algorithm to determine the coreness of a node based on local information. Approximating the coreness/shell number locally may improve the time complexity of the proposed algorithms.

\bibliographystyle{IEEEtran}
\bibliography{bibfile}

\newpage
\appendix

\section{APPENDIX}

\subsection{Datasets Overview}

Table \ref{datasets} gives a short description of the all the datasets used in this paper.
\begin{table}
\begin{center}
\begin{tabular}{ |m{2cm}|m{10cm}| }
\hline
\begin{center}\scriptsize{\textbf{Dataset}}\end{center}& \begin{center}\scriptsize{\textbf{Description}}\end{center} \\
\hline
\scriptsize{Facebook} & \scriptsize{Facebook is the most popular Social Networking Site today. This dataset consists of anonymized friendship relations from Facebook. \cite{leskovec2012learning}. The network contains 4,039 nodes and 88,234 edges.}\\
\hline
\scriptsize{Google plus} & \scriptsize{Google plus is a social layer for Google Services.\cite{leskovec2012learning}. The network contains 107,614 nodes and 13,673,453 edges.}  \\
\hline
\scriptsize{Slashdot} & \scriptsize{Slashdot is a website where the users can submit and evaluate the news stories on science and technology. It is famous for its  specific user community. \cite{leskovec2009community}. This dataset contains 82,168 nodes and 948,464 edges.}\\
\hline
\scriptsize{Flickr} & \scriptsize{This is an image and video hosting site. It is mainly used for sharing and embedding personal photographs. \cite{Zafarani+Liu:2009}. This dataset has 80513 nodes and 5899882 edges.}
 \\
\hline
\scriptsize{Livemocha} & \scriptsize{Livemocha is the world's largest online language learning community.\cite{Zafarani+Liu:2009}. This dataset has 104438 nodes and 2196188 edges.} \\
\hline
\scriptsize{DBLP} & \scriptsize{The DBLP computer science bibliography is a collaboration network. It provides a detailed list of research papers in computer science \cite{yang2015defining}. The network contains 317,080 nodes and 1,049,866 edges.} \\
\hline
\scriptsize{Buzznet} & \scriptsize{Buzznet is social media network used for sharing  photos, journals, and videos. It has 101168 nodes and 4284534 edges.} \\
\hline
\end{tabular}\\
\vspace{2mm}
\caption{Datasets used for experiments}
\label{datasets}
\end{center}
\end{table}

\subsection{Leakage Power}
This definition formally defines the leakage power introduced in the future section of the paper.
For defining the leakage power, first we define a ``Teleportation Edge'':\\

\textbf{Teleportation Edge}: An edge $E_{ij}$ in the network is called a teleportation edge if $shell(j)>shell(i)$. Let T be the set of teleportation edges i.e. $E_{ij} \in T\ iff shell(j)>shell(i)$\\
The leakage power of a shell $S$ is proportional to the number and the average height of the ladders present in this shell.
Let the vertex set of a shell $S$ be represented as $V(S)$ and edge set as $E(S)$. Let the leakage power of this shell be represented by $\theta_{S}$.

Then ,

$\theta_{S} \propto |E_{ij}|$ where $i \in V(S)$ $\wedge E_{ij} \in T$.

Let the height of a teleportation edge $E_{ij} \in T$
be defined by $H_{ij} = shell(j)-shell(i)$

Then,

$\theta_{S} \propto \frac {\sum_{e_{ij} \in E(S)} H_{ij}}{m \times (n-m) }$. \hspace{1cm} where $m=|V(S)|$ and $n=|V(G)|$ $\wedge E_{ij} \in T$\\
 
 So, overall
 
 $\theta_{S} = \kappa |E_{kl}| \frac {\sum_{e_{ij} \in E(S)} H_{ij}}{m \times (n-m) }$. \hspace{1cm} where $k \in V(S)$, $m=|V(S)|$, $n=|V(G)|$, and $E_{kl} \in T$. \\
 
 $\kappa$ is a parameter whose value depends on the network and needs to be suitably found out.

\subsection{Results of the proposed algorithms}

\begin{figure}[h!]
    \centering
    \subfigure[Random Walk]
    {
       \includegraphics[width=9 cm]{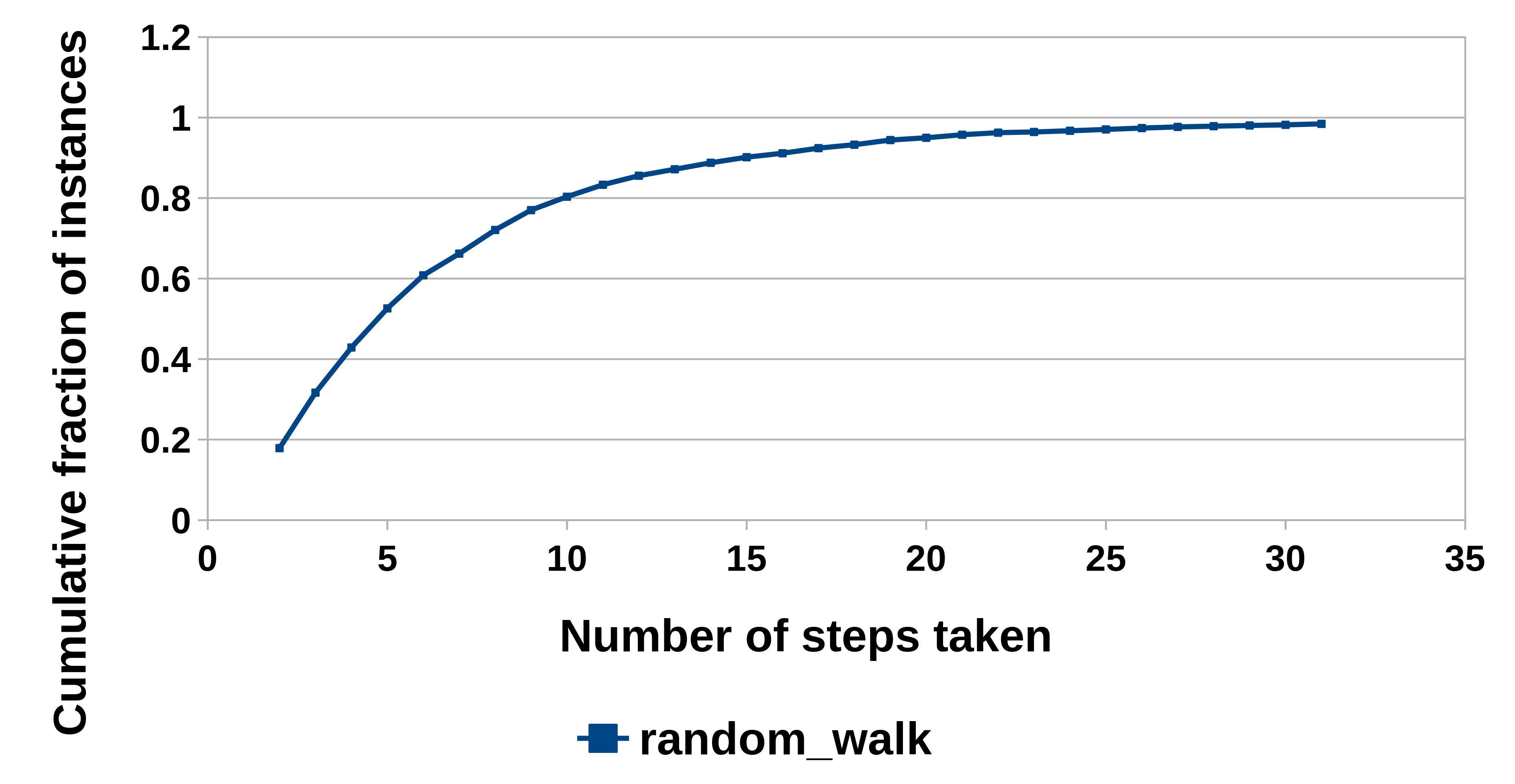}
       
    }
    \subfigure[Shell Based Hill Climbing Algorithms]
    {
       \includegraphics[width=9 cm]{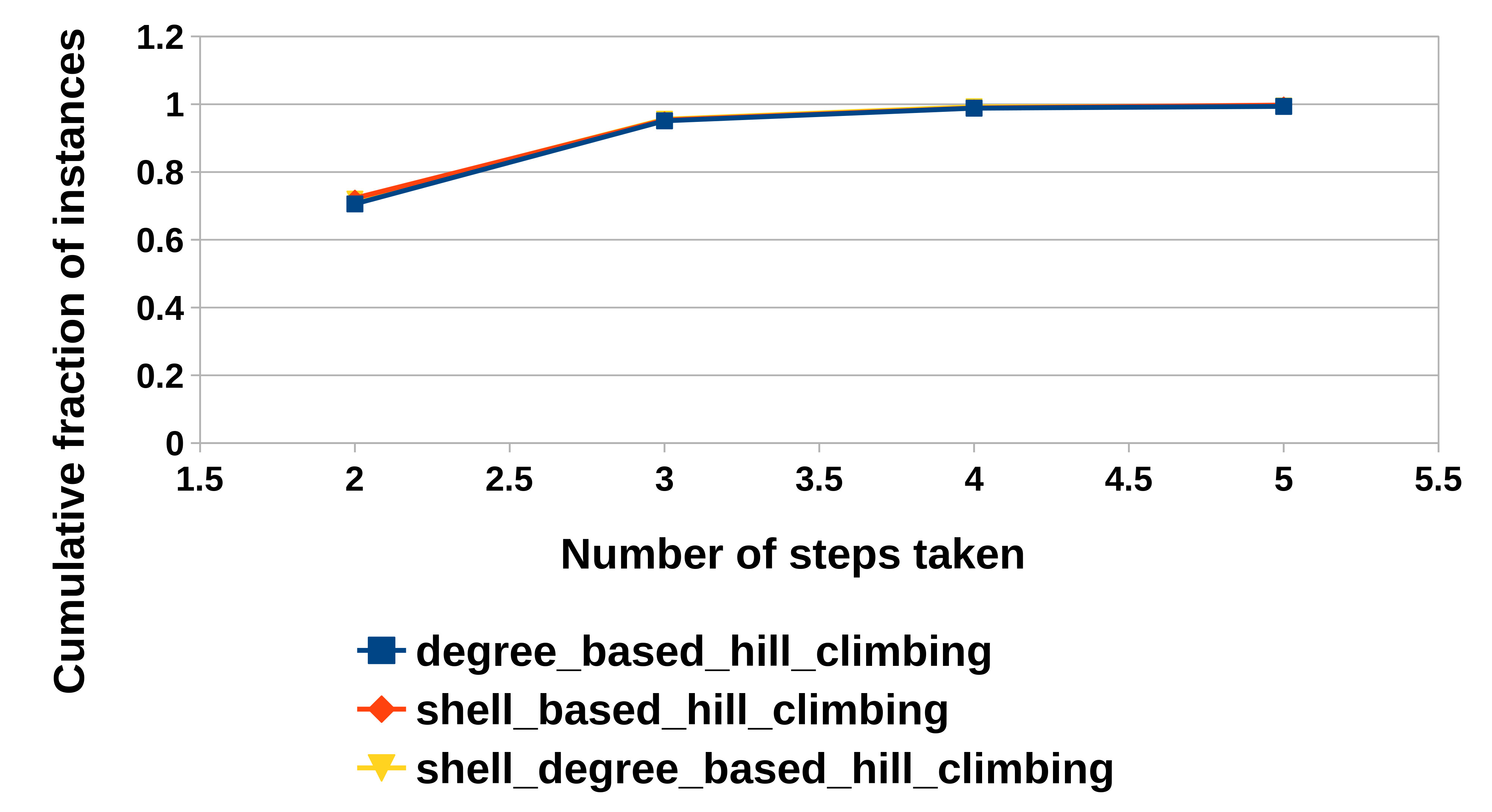}
       
    }
  \caption{Comparison of algorithms for Slashdot}
  \label{sd}
\end{figure}

\begin{figure}[h!]
    \centering
    \subfigure[Random Walk]
    {
       \includegraphics[width=9 cm]{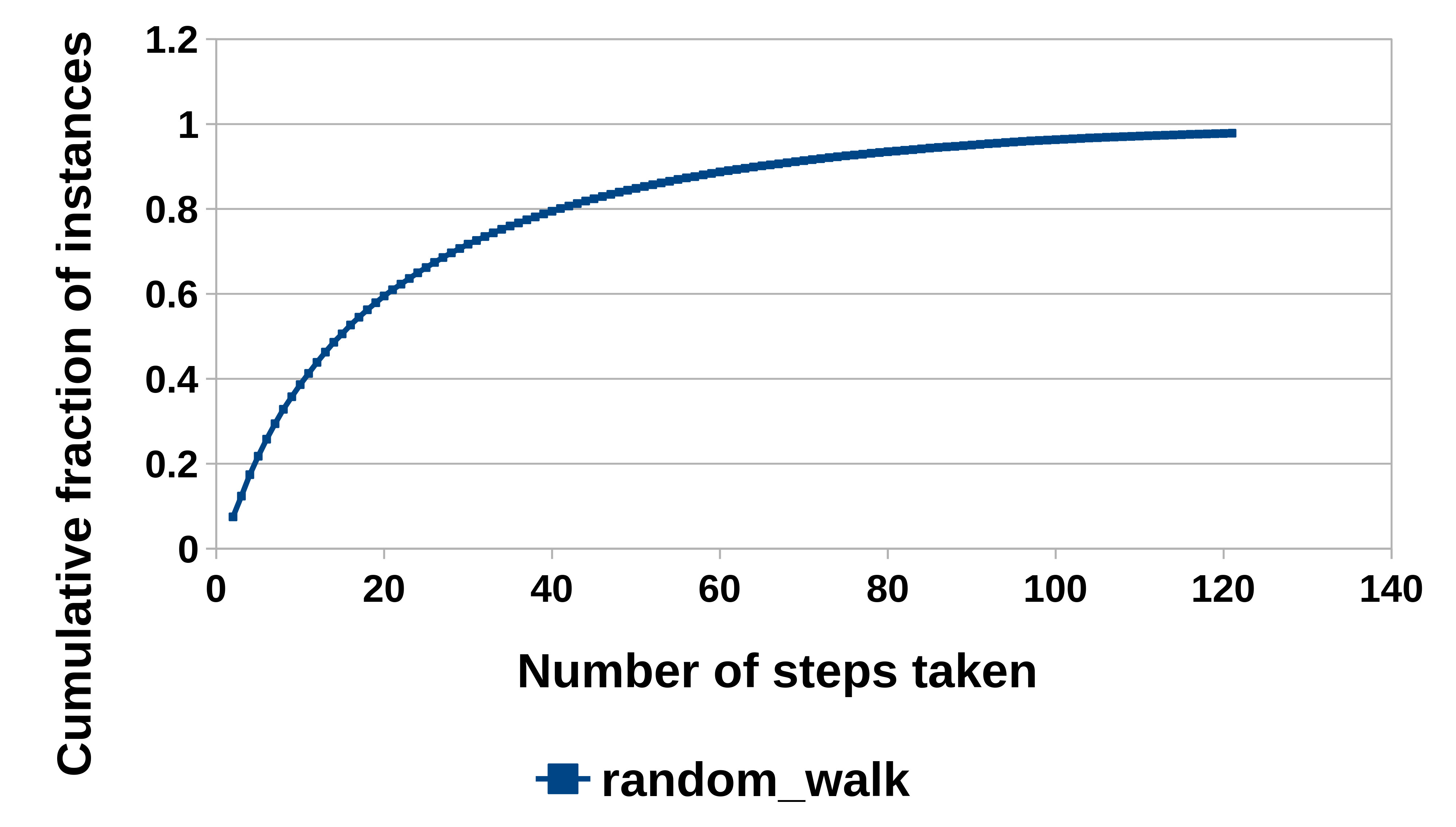}
       
    }
    \subfigure[Shell Based Hill Climbing Algorithms]
    {
       \includegraphics[width=9 cm]{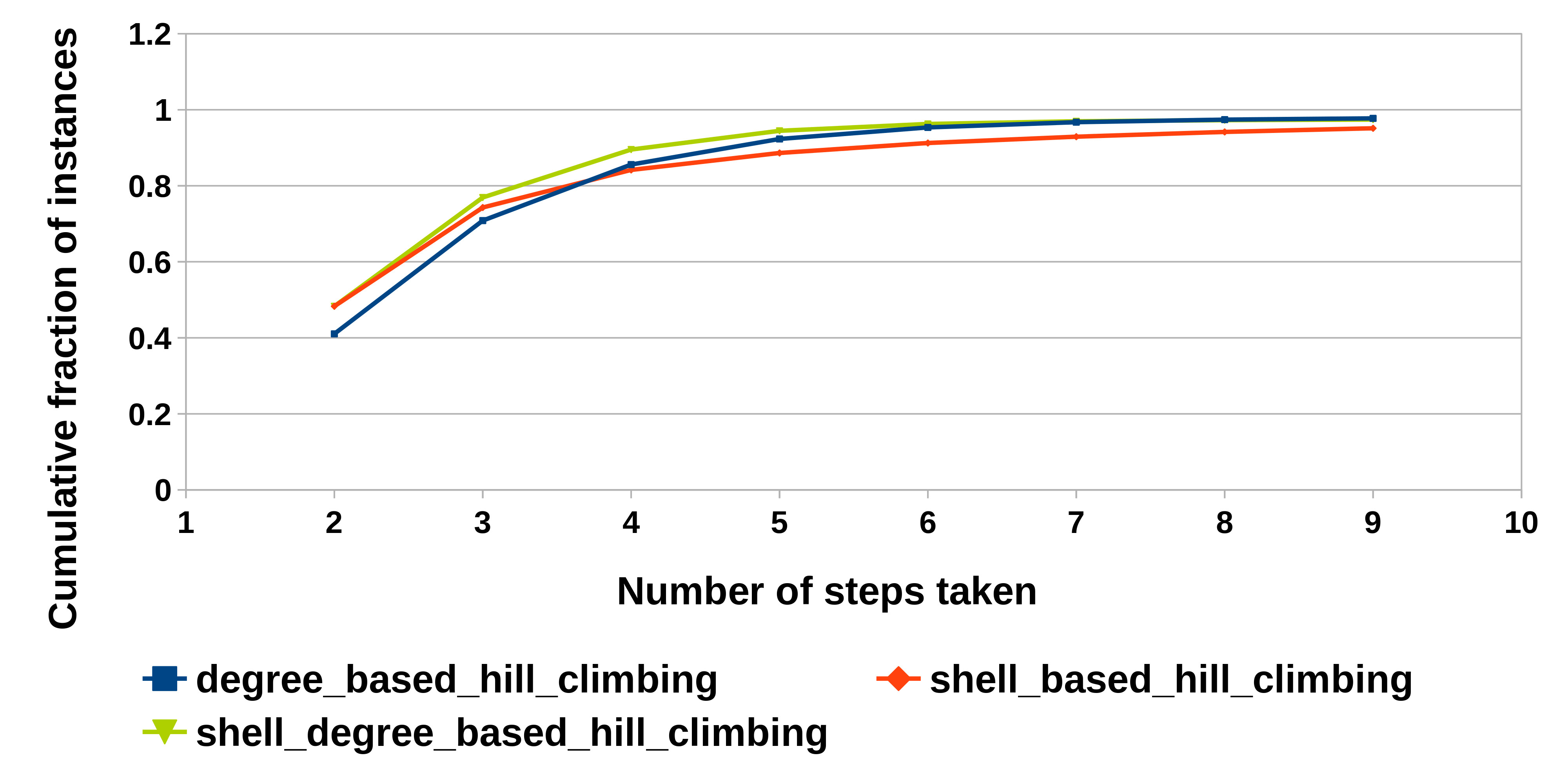}
       
    }
  \caption{Comparison of algorithms for DBLP}
  \label{db}
\end{figure}

\begin{figure}[h!]
    \centering
    \subfigure[Random Walk]
    {
       \includegraphics[width=9 cm]{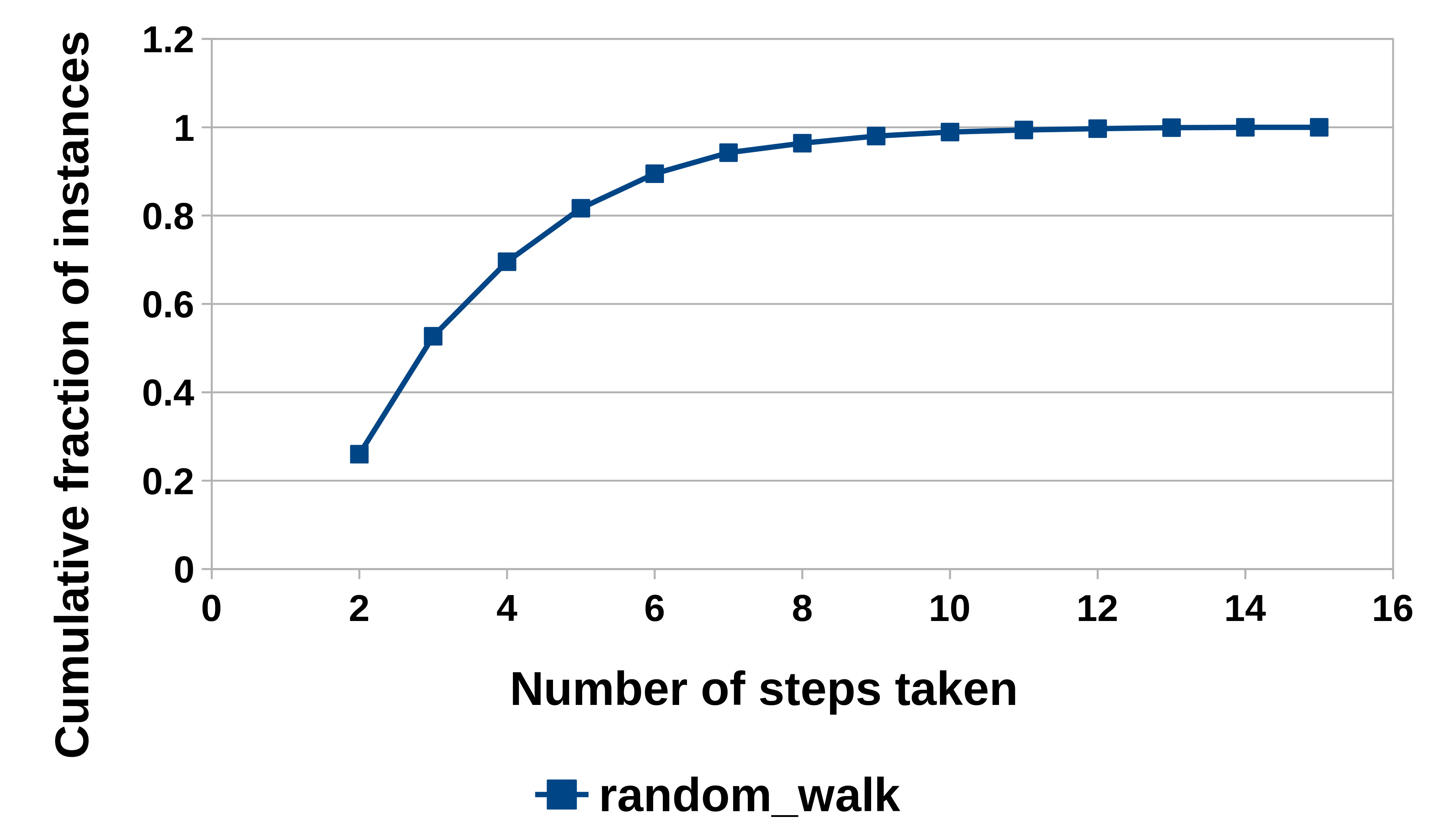}
       
    }
    \subfigure[Shell Based Hill Climbing Algorithms]
    {
       \includegraphics[width=9 cm]{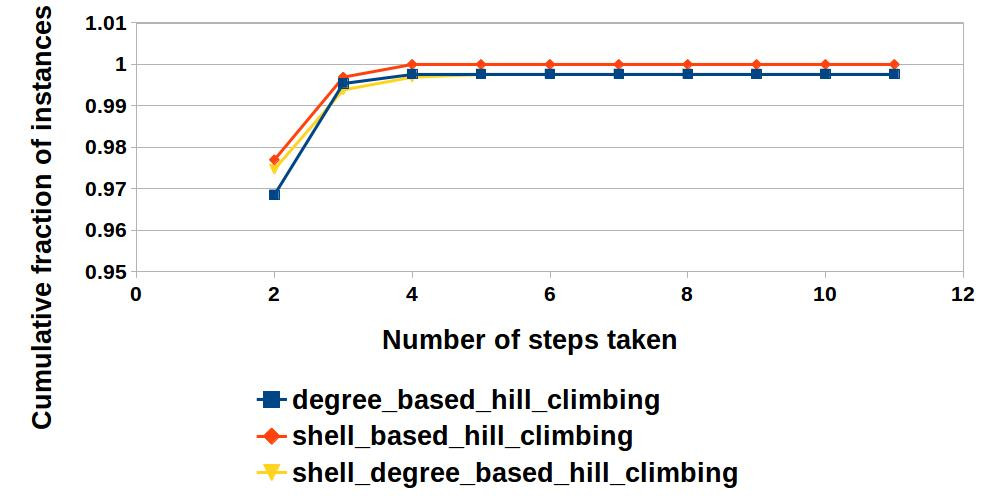}
       
    }
  \caption{Comparison of algorithms for Livemocha}
  \label{lm}
\end{figure}

The figures \ref{sd}, \ref{db} and \ref{lm} show the results of the algorithms discussed in the paper for three networks for the case when the destination is a core shell.

\end{document}